\documentclass[fleqn,usenatbib]{mnras}
\usepackage{float}
\usepackage{amsmath}
\usepackage{siunitx}

\usepackage{newtxtext,newtxmath}

\usepackage[T1]{fontenc}
\usepackage[normalem]{ulem} 
\DeclareRobustCommand{\VAN}[3]{#2}
\let\VANthebibliography\thebibliography
\def\thebibliography{\DeclareRobustCommand{\VAN}[3]{##3}\VANthebibliography}

\usepackage{graphicx}	
\usepackage{amsmath}	
\usepackage{orcidlink}

\title[\texttt{SPICE} and the UVLF variability]{Variability of the UV luminosity function with \texttt{SPICE}}

\author[Basu et al.]{
Arghyadeep Basu \orcidlink{0000-0001-8104-9751} $^{1,2,3}$\thanks{E-mail: basu@mpa-garching.mpg.de}\thanks{E-mail: basu.arghyadeep@yahoo.in},
Aniket Bhagwat \orcidlink{0000-0003-0275-5506}$^{1}$,
Benedetta Ciardi \orcidlink{0000-0002-5037-310X} $^{1}$ \&
Tiago Costa \orcidlink{0000-0002-6748-2900}$^{4}$
\\
$^{1}$Max-planck-Institut f$\ddot{u}$r Astrophysik, Karl-Schwarzschild-Strasse 1, D-85741, Garching, Germany\\
$^{2}$Ludwig-Maximilians-Universität München (LMU), Geschwister-Scholl-Platz 1, 80539 München, Germany\\
$^{3}$Univ Lyon, Univ Lyon1, Ens de Lyon, CNRS, CRAL UMR5574, F-69230, Saint-Genis-Laval, France\\
$^{4}$School of Mathematics, Statistics and Physics, Newcastle University, NE1 7RU, UK \\
}

\date{Accepted 2025 October 1. Received 2025 September 24; in original form 2025 January 30}

\pubyear{2015}

\begin{document}
\label{firstpage}
\pagerange{\pageref{firstpage}--\pageref{lastpage}}
\maketitle

\begin{abstract} 
We investigate the variability of the UV luminosity function (UVLF) at $z > 5$ using the \texttt{SPICE} suite of cosmological, radiation-hydrodynamic simulations, which include three distinct supernova (SN) feedback models: \texttt{bursty-sn}, \texttt{smooth-sn}, and \texttt{hyper-sn}. 
The \texttt{bursty-sn} model, driven by intense and episodic SN explosions, produces the highest fluctuations in the star formation rate (SFR). Conversely, the \texttt{smooth-sn} model, characterized by gentler SN feedback, results in minimal SFR variability. The \texttt{hyper-sn} model, featuring a more realistic prescription that incorporates hypernova (HN) explosions, exhibits intermediate variability, closely aligning with the \texttt{smooth-sn} trend at lower redshifts.
These fluctuations in SFR significantly affect the $\rm{\textit{M}_{UV} - \textit{M}_{halo}}$ relation, a proxy for UVLF variability. Among the models, \texttt{bursty-sn} produces the highest UVLF variability, with a maximum value of 2.5. In contrast, the \texttt{smooth-sn} and \texttt{hyper-sn} models show substantially lower variability, with maximum values of 1.3 and 1.5, respectively.
However, in all cases, UVLF variability strongly correlates with host halo mass, with lower-mass halos showing greater variability due to more effective SN feedback in their shallower gravitational wells. The \texttt{bursty-sn} model, though, results in higher amplitudes.
Variability decreases in lower mass haloes with decreasing redshift for all feedback models.
This study underscores the critical role of SN feedback in shaping the UVLF, and highlights the mass and redshift dependence of its variability, suggesting that UVLF variability may alleviate the bright galaxy tension observed by \texttt{JWST} at high redshifts.

\end{abstract}




\section{Introduction}

Since the James Webb Space Telescope (\texttt{JWST}) started to provide revolutionary data in July 2022, our understanding of the early galaxy formation process has been challenged.  
Although over the past decade the Hubble Space Telescope (\texttt{HST}) enabled us to observe galaxies out to $z \approx 10$ \citep{zheng2012,coe2013,oesch2016,morishita@ARTICLE,bagley2022},  \texttt{JWST} has pushed our observational barrier to even higher redshifts, and has detected an unexpected overabundance of massive galaxies at $z > 10$, exceeding theoretical predictions that anticipate much lower number densities due to the limited time available for galaxy formation and growth at such early epochs \citep{castellano2022,finkelstein2022,naidu2022,adams2023,morishita_stiavelli2023,bouwens2023a,bouwens2023b,donnan2023,atek2023,perez2023,willott2024,Donnan2024}. 
This discovery might challenge theoretical models, as it suggests a high incidence of bright galaxies at this epoch
\citep{mason2023,boylan2023,lovell2023,acharya2024}.
The recent detection of a spectroscopically confirmed galaxy at $z=14.32$ by \citet{helton2024} significantly elevates the challenge, suggesting that such bright objects exist even at these high redshifts. These observations raise several intriguing questions: Could there be a missing population of bright stars? Is there something unique about the early star formation history? Do these findings create tensions with the standard $\Lambda$CDM cosmology? 

A few potential physical processes have been suggested to explain the discovery of these bright galaxies. One of them is an increased star formation efficiency (SFE) due to a weaker feedback which could boost the abundance of UV-bright galaxies by forming more stars per baryon \citep{fukushima2022,inayoshi2022,harikane2023}. Also, a top-heavy initial mass function (IMF) could lead to an abundance of early bright galaxies by producing more UV photons per unit of stellar mass formed \citep{inayoshi2022,yung2023,chon2024,steinhardt2023,Cueto2024,hutter2024}. The interplay between dust attenuation and the abundance of massive halos at high redshifts could lead to a mild evolution of the UV luminosity function (LF). While this alone cannot fully explain the abundance of high-redshift bright galaxies \citep{ferrera2023,mirocha_furlanetto2023}, radiation pressure on dust may contribute to the reduction of the attenuation \citep{Fiore2023,ferrara2024b}. 
More exotic scenarios invoked include a modified primordial power spectrum \citep{hirano_yoshida2023,padmanabhan2023,parashari2023,sabti2023}, primordial non-Gaussianity \citep{biagetti2023}, and alternative dark matter scenarios \citep{bird2023,dayal_giri2023,gong2023}.
In this respect, though, it is important to acknowledge several key factors that may influence these observations. For instance, spectral energy distribution (SED) fitting methods often carry significant uncertainties \citep{lower2020,hollis2023,jain2024,haskell2024}. Additionally, observational biases, the potential misinterpretation of field overdensities, and challenges in making proper comparisons to simulations may further complicate the understanding of early bright galaxy populations \citep{narayanan2024,luberto2024,nguyen2024,whitler2024}
.

Another topic that has attracted significant attention is the variability of the UVLF of early galaxies. 
A possible explanation for its origin is a strong stochasticity of star formation. Indeed, both simulations and observations indicate that small dwarf galaxies and high-redshift galaxies display stochastic star formation histories \citep{sparre2017,smit2016,emami2019,iyer2020,tachhella2020,flores2021,hopkins2023},
which are also associated to irregular and clumpy morphologies of the hosting galaxies \citep{bournard2007,elmegreen2009,forster2011,true2023}. The variability in star formation efficiency is likely driven by a combination of factors, including gas inflow and outflow dynamics, gravitational instabilities, and galaxy mergers during the early stages of galaxy formation \citep{dekel2009,ceverino2010,angles2017}, intense feedback \citep{elbadry2016,tacchella2016} or sometimes even feedback-free starbursts \citep{faucher2018,Dekel2023}. These combined effects alter the galactic environment, leading to variability in the UVLF \citep{sparre2017,furlanetto2022,sun2023}. 

In this paper, we investigate the impact of SFR variability on the UVLF of high-$z$ galaxies using the suite of radiation hydrodynamic simulations \texttt{SPICE} \cite[][hereafter B24]{Bhagwat2024}, which comprises three different implementations of supernova (SN) feedback (while the rest of the simulation setup is the same), resulting in a different level of burstiness and  in a variety of star formation histories. This allows us to systematically quantify the SFR and UVLF variability.

Specifically, we introduce the simulations  in Section \ref{section:2}, the results are presented in Section \ref{section:3}, while in Section \ref{section:4} we summarise our conclusions and future prospects. Throughout the paper, we adopt a flat $\Lambda \rm{CDM}$ cosmology consistent with \cite{Planck2016} with $\Omega_{\rm m}=0.3099$, $\Omega_{\Lambda}=0.6901$, $\Omega_{\rm b}=0.0489$, $h=0.6774$, $\sigma_{8}=0.8159$ and $n_{\rm s}=0.9682$, where the symbols have their usual meaning.

\section{The \texttt{SPICE} simulation suite}
\label{section:2}
In order to investigate the impact of SN feedback onto galaxy properties, we post-process the suite of radiation-hydrodynamical simulations \texttt{SPICE} (B24), which we briefly introduce below.

\texttt{SPICE} has been performed with \texttt{RAMSES-RT} \citep{rosdahl2013,rosdahl2015}, the radiation-hydrodynamics extension of the Eulerian Adaptive Mesh Refinement (AMR) code \texttt{RAMSES} \citep{teyssier2002}. The simulations target a cubic box of size 10 $\rm{\mathit{h}^{-1} cMpc}$ with $512^{3}$ dark matter particles of mean mass $m\rm{_{dm}} = 6.38 \times 10^{5} \mathrm{M_{\odot}}$. The AMR allows each parent gas cell to split into 8 smaller cells when certain conditions are met (see B24). 
As a reference, this provides a physical resolution at $z=5$ ranging from 4.8 kpc for the coarsest level ($\ell_{\text{min}} = 9$) to approximately 28 pc for the finest level ($\ell_{\text{max}} = 16$). 
The hydrodynamical equations are solved using a second order Godonov scheme and the dynamics of collisionless dark matter and stellar particles are computed by solving the Poisson equation with a particle-mesh solver, projecting onto a grid using the Cloud-in-Cell  scheme \citep{guilet2011}. Gas cooling and heating are accounted for as in \citet{rosdahl2013}. Star-formation has been modelled following \citet{kretschmer2020}, which adopt a subgrid turbulence model depending on the physical conditions of each computational cell including parameters like the local gas turbulent Mach number and the virial parameter, which acts as a measure of the gravitational balance in the region, and which results in a spatially-variable star formation efficiency. Additionally, \texttt{SPICE} includes a radiative transfer scheme with five frequency groups, i.e. infrared (0.1-1 $\rm{eV}$), optical (1-13.6 $\rm{eV}$), and three UV (13.6-24.59 $\rm{eV}$; 24.59-54.42 $\rm{eV}$; 54.42-$\infty$ $\rm{eV}$).

To correctly capture the momentum injection into the cells due to supernova feedback, the scheme presented in \citet{kimm2015} is adopted. Keeping all the parameters for momentum injection constant, different simulations have been run based on the timing of supernova events and the amount of energy injected. We briefly present the different feedback models as follows:
\begin{itemize}
    \item "\textbf{\texttt{bursty-sn}}": each stellar particle experiences a single event of SN explosions at a fixed timescale of 10 Myr, equivalent to the mean time at which SN occurs for the adopted \citet{chabrier2003} initial mass function. An energy of $2 \times 10^{51}$ ergs per SN explosion is injected into the neighbouring cells.
    \item "\textbf{\texttt{smooth-sn}}": for a given stellar population, a SN with mass $> 20 \rm{M_{\odot}}$ can explode as early as 3 Myr, whereas SN with progenitor mass of $8 \rm{M_{\odot}}$ can explode as late as 40 Myr. An energy of $2 \times 10^{51}$ ergs per SN explosion is injected into the neighbouring cells.
    \item "\textbf{\texttt{hyper-sn}}": SN explosions happen at different times as in \texttt{smooth-sn}, with the injected energy extracted stochastically from a normal distribution centered at $1.2 \times 10^{51}$~ergs and ranging from $10^{50}$~ergs to $2 \times 10^{51}$~ergs (see also, \citealt{sukhbold2016,diaz2018,diaz2021}). Additionally, a metallicity dependent fraction of SN explodes as Hypernovae (HN), with an energy injection of $10^{52}$ ergs.
\end{itemize}

Dust in \texttt{SPICE} is not modelled as an active ingredient but rather through a direct dependence on the local gas-phase properties, i.e. its metallicity and ionization state. Following \citet{Nickerson2018}, the dust number density is computed as
$n_{\rm d} = (Z/Z_\odot)  f_{\rm d} n_{\rm H}$, 
where $ Z $ represents the local metallicity, $ f_{\rm{d}} = 1 - x_{\rm{HII}} $ is the fraction of neutral hydrogen, and $ n\rm{_{H}} $ is the hydrogen number density. 
This relation ensures that the dust density depends on both chemical enrichment and the local ionization conditions.
Indeed, despite expectations of dust being typically destroyed in photo-ionized gas, observations have shown dust-to-ionized-gas ratios as high as 0.01 \citep{harper1971,contini2003,laursen2009}. Thus, the quantity $f_{\rm{d}}$ permits non-zero dust effects even in partially ionized environments.
The propagation of UV photons influences the gas via photo-ionization and photo-heating, but also through radiation pressure generated from both photo-ionization and through dust absorption. Infrared and optical photons interact with gas solely through radiation pressure on dust particles. 
Each photon group is characterized by specific dust absorption and scattering properties. When optical and UV photons are absorbed by dust, their energy is transferred to the infrared group, which later interacts with the gas through multi-scattering radiation pressure.

For a more comprehensive description of the simulations and the supernova feedback models discussed above, we refer the readers to B24. 

\section{Results}
\label{section:3}
In this Section, we present results from our analysis of how different SN feedback models affect the SFR and UVLF. 

\subsection{UV luminosity function}
\label{section:3.1}
\begin{figure}
    \includegraphics[width=\columnwidth]{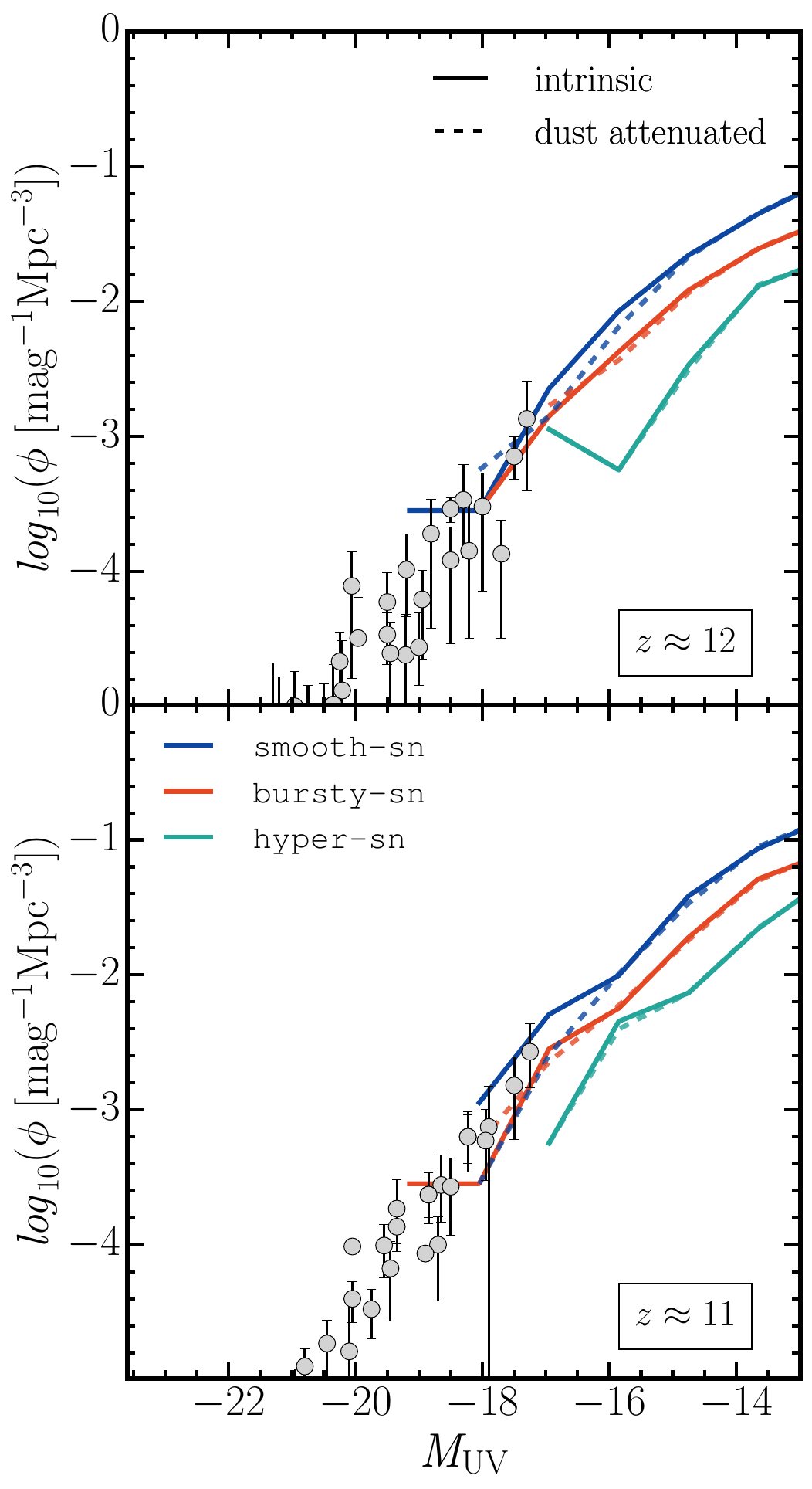}
  \caption{1500~$\si{\angstrom}$ luminosity functions at $z$ = 12 (\textit{top panel}) and 11 (\textit{bottom panel}) for three feedback models (in different colors). Solid and dashed curves refer to intrinsic and dust attenuated LFs, respectively. A compilation of observations from \texttt{HST} and \texttt{JWST} \citep{bouwens2015,harikane2022,naidu2022,adams2023,harikane2023,bouwens2023a,bouwens2023b,leung2023,donnan2023a,donnan2023b,perez2023,casey2024,robertson2024,mcleod2024,Whitler2025} is shown as gray data points.}
  \label{fig:uvlf}
\end{figure}

While in B24 we showed results at $z\leq10$, in Figure \ref{fig:uvlf} we present the 1500~\AA\ LF for all feedback models  at $z=11$ and 12. 
The LFs have been derived from the stellar spectral energy distribution within a 10~$\si{\angstrom}$ bin centered around 1500~$\si{\angstrom}$ . We use the SED model of BPASSv2.2.1 \citep{elridge2017,stanway2018}, assuming the \citet{chabrier2003} IMF (details are mentioned in B24). The solid and dashed lines in the figure indicate the intrinsic and dust-attenuated luminosity functions, respectively. 
To account for dust attenuation, we utilize the Monte Carlo radiative transfer code \texttt{RASCAS} \citep{michel2020}. In this approach, 100 rays are cast from each stellar particle within a halo out to its virial radius. The dust attenuation along each ray is computed following the model described in the previous section.
Note that, rather than applying a fixed star formation rate and UV luminosity conversion, the BPASSv2.2.1 SED model treats these two quantities independently while preserving their intrinsic correlation, thereby allowing us to capture their natural variability.

As already noted in B24, also at these redshifts \texttt{smooth-sn} produces a higher intrinsic luminosity function (by 0.3-0.4 dex) at all magnitudes. 
Conversely, the \texttt{hyper-sn} model yields the lowest galaxy count across all magnitude bins. 
The \texttt{bursty-sn} model produces results which are typically in between those of the other models. 
The effect of dust attenuation (dashed curves) starts to be visible from an absolute magnitude of $M_{\rm{UV}} \approx -15$, the value changing slightly depending on feedback model and redshift. 
When comparing the dust attenuated LFs to a compilation of observations from \texttt{HST} and \texttt{JWST} \citep{bouwens2015,harikane2022,naidu2022,adams2023,harikane2023,bouwens2023a,bouwens2023b,leung2023,donnan2023a,donnan2023b,perez2023,casey2024,robertson2024,mcleod2024,Whitler2025}, we notice a fairly good match in the range of magnitudes covered by \texttt{SPICE} (except for the \texttt{hyper-sn} model), which, given the limited box size, does not extend to the brightest observed galaxies. 
For a more detailed discussion of the LF, we refer the reader to B24. 

\begin{figure}
        \includegraphics[width=\columnwidth]{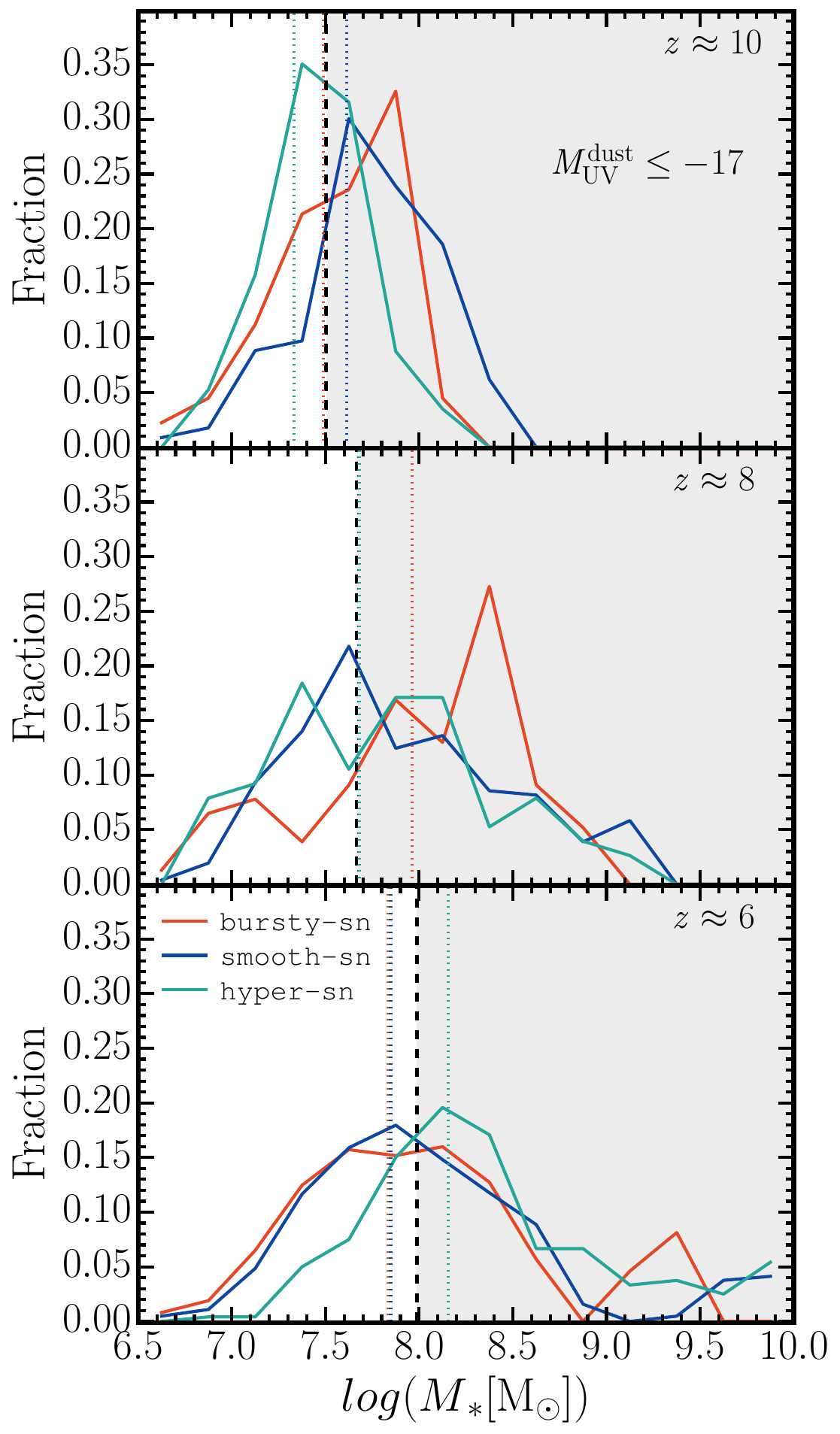}
  \caption{Distribution of stellar mass of objects with $\rm{\mathit{M}_{UV}^{dust}} \leq -17$ at  $z=10$ (\textit{top panel}), 8 (\textit{middle}) and 6 (\textit{bottom}). The corresponding median stellar masses are shown as vertical dotted lines, while the black dashed lines refer to the minimum stellar mass required for having such bright objects assuming the median SFE model of \citet{mason2015,mason2023}.}
  \label{fig:hoststellarmass_dist}
\end{figure}

In Figure \ref{fig:hoststellarmass_dist}, we further investigate the contribution to the bright end of the UVLF (with $\rm{\mathit{M}_{UV}^{dust}} \leq -17$) from objects with different stellar masses. To obtain a statistical sample, we have derived the distributions using all the simulation snapshots within a time interval of 100 Myr centered on each redshift shown. We observe that, for all feedback models, the stellar masses are always spread over a wide range, which extends to increasingly larger masses as redshift decreases, indicating that even objects with stellar masses as low as $10^{(6.5-7)} \rm{M_{\odot}}$ can contribute to the bright end of the LF. For all models the peak of the distribution drops with time along with the distribution becoming flatter and broader. Notably, though, all models exhibit a similar distribution, although the corresponding medians are  slightly different,  
with \texttt{smooth-sn}, \texttt{bursty-sn} and \texttt{hyper-sn} having the largest median value at $z\approx10$, 8 and 6, respectively. 
For a comparison to previous studies, we also show the stellar mass threshold required for having objects with $\rm{\mathit{M}_{UV}^{dust}} \leq -17$, obtained in the semi-analytic model of \citet{mason2015,mason2023} (assuming the median SFE model), which has been used to explain the abundance of bright galaxies  detected by \texttt{JWST} at $z>10$.  
It is evident that their estimates are very similar to the median values obtained in our simulations. 

\begin{figure*}
    \includegraphics[width=176mm]{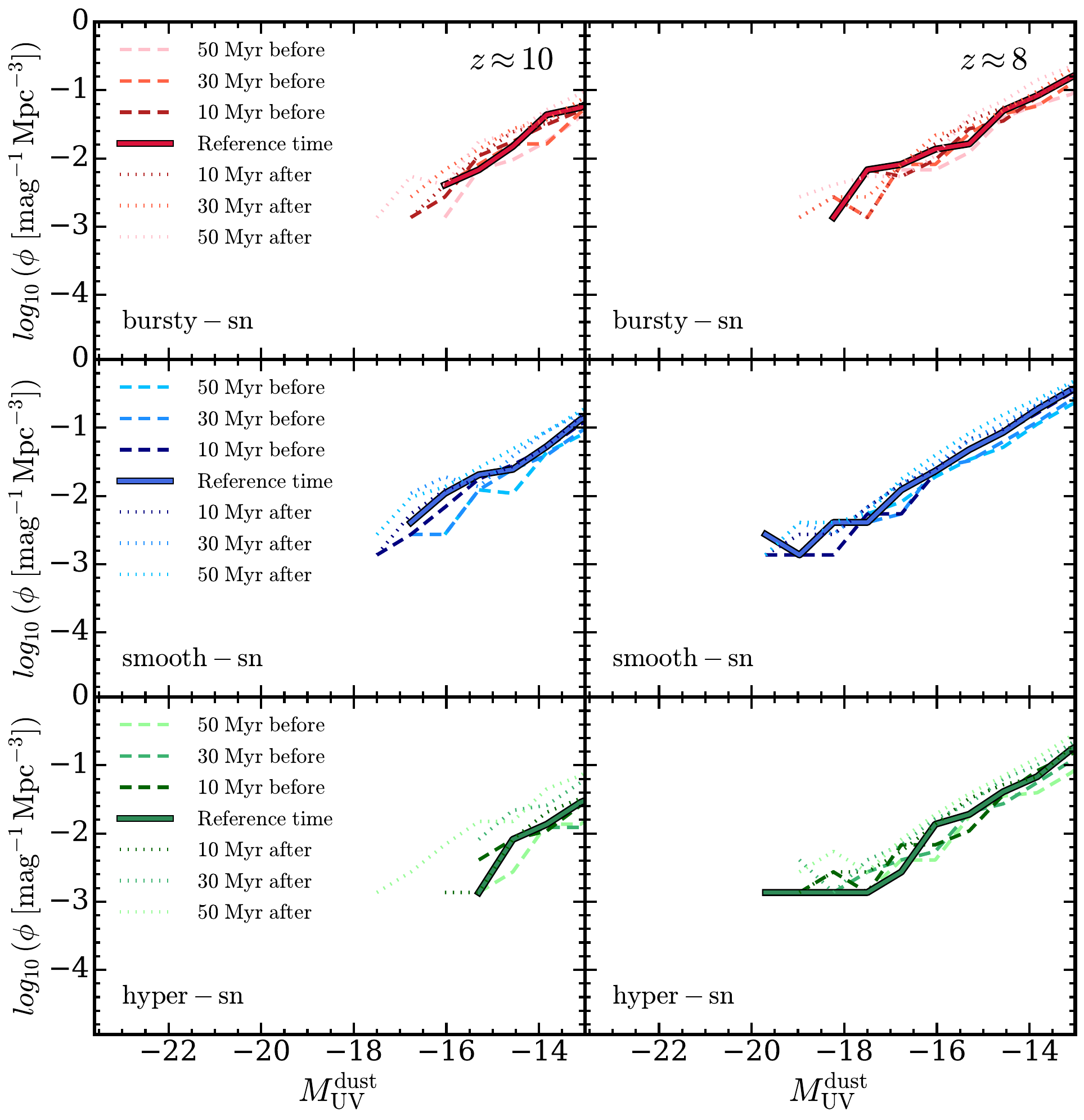}
  \caption{
Evolution of the dust-attenuated UVLF over time intervals of 10, 30, and 50 Myr (indicated by different color gradients, from dark to light), centered at $z \approx 10$ (\textit{left panel}) and 8 (\textit{right}), for three different feedback models: \texttt{bursty-sn} (\textit{top panels}), \texttt{smooth-sn} ({\it middle}), and \texttt{hyper-sn} ({\it bottom}). The dashed and dotted curves show the UVLF before and after the reference time.
}
  \label{fig:uvlf_temporal_evol}
\end{figure*}

\begin{figure*}
        \includegraphics[width=175mm]{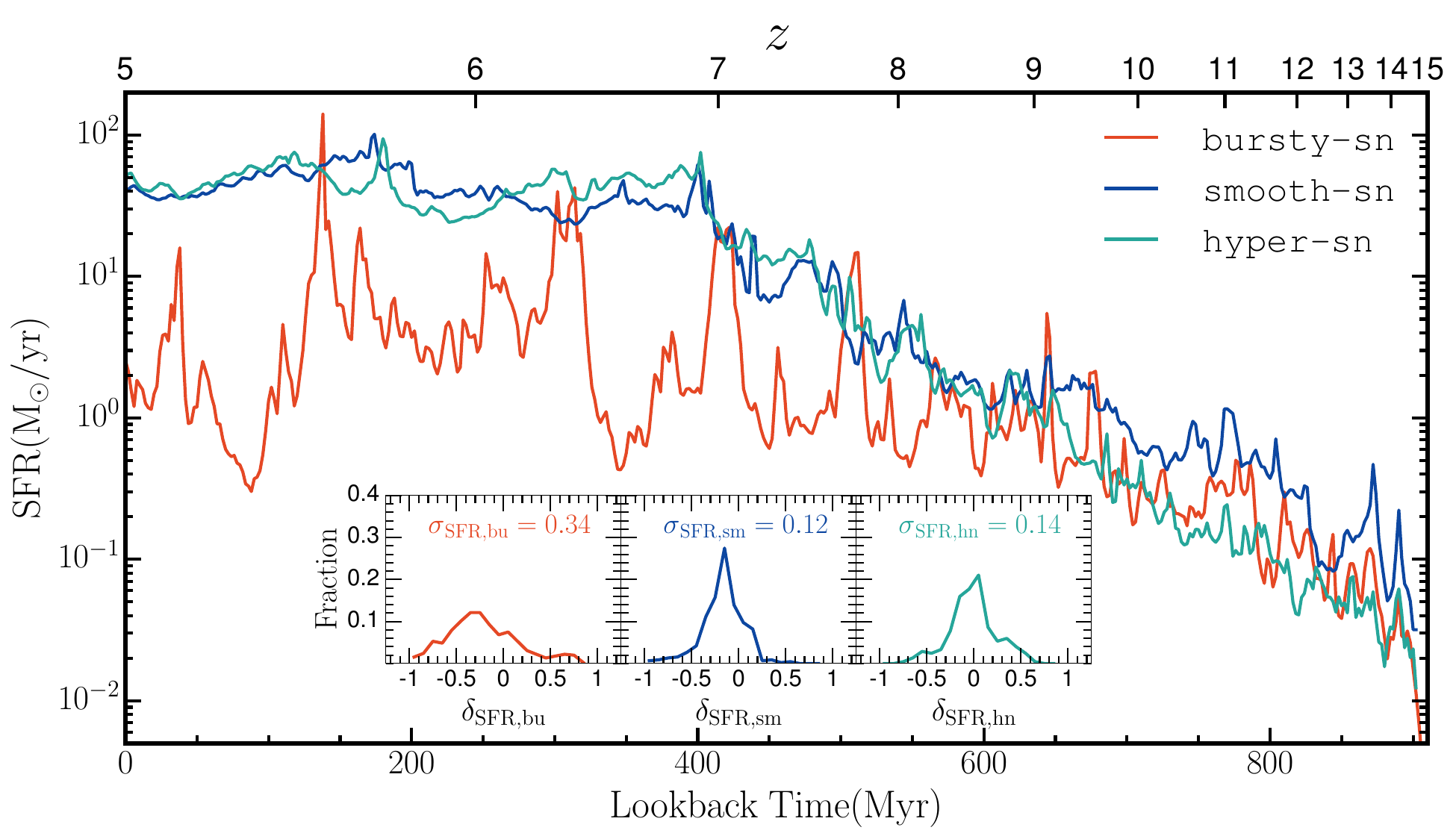}
  \caption{Temporal evolution of the SFR averaged over 2 Myr accounting for all stellar particles inside the virial radius of the most massive halo at $z = 5$, which has a total stellar mass of $2.1 \times 10^{9} \rm{\mathrm{M_{\odot}}}$, $1.9 \times 10^{10} \rm{\mathrm{M_{\odot}}}$ and $1.7 \times 10^{10} \rm{\mathrm{M_{\odot}}}$ for the \texttt{bursty-sn}, \texttt{smooth-sn} and \texttt{hyper-sn} model, respectively. Colors refer to the different SN feedback models. In the inset, we show the distributions of $\delta\rm{_{SFR,X}}$, together with the corresponding standard deviation, $\sigma\rm{_{SFR,X}}$ (see text for the details of the calculation),where `X' = `bu', `sm', `hn' for the \texttt{bursty-sn}, \texttt{smooth-sn} and \texttt{hyper-sn} model, respectively.
  }
  \label{fig:sfr_individual}

\end{figure*}

To investigate the temporal evolution of the dust-attenuated UVLF on short timescales, in Figure~\ref{fig:uvlf_temporal_evol} we present the UVLF in intervals of 10, 30, and 50 Myr around $z \approx 10$ and 8. While the general behaviour in the three models is qualitatively similar, important differences emerge in the nature of the UVLF variability. The UVLFs at these epochs are shaped by a combination of population growth, feedback-driven luminosity variability, dust attenuation, and the dominance of a few bright systems.
At $z \approx 10$, population growth is the dominant effect, producing a consistent rise across all magnitude bins as we probe the very early stages of galaxy evolution in the simulation volume. This growth-driven behaviour is most evident in the \texttt{smooth-sn} and \texttt{hyper-sn} models. In the \texttt{bursty-sn} model, however, the SN feedback driven stochasticity of the star formation rate and thus of UV luminosity (we discuss this aspect later) plays comparable role, reducing the coherence of the uniform rise. Overall, at this redshift the apparent UVLF variability primarily reflects the rapid evolution of the galaxy population, rather than true short-timescale fluctuations of the UVLF.
By $z \approx 8$, feedback-driven luminosity variability becomes more pronounced, with the \texttt{bursty-sn} model exhibiting the strongest fluctuations. In comparison, the \texttt{smooth-sn} model again shows a more gradual, monotonic evolution in the number of galaxies, consistent with the gentler nature of its SN feedback. The behaviour of the \texttt{hyper-sn} model lies in between: a combination of population growth and genuine feedback-driven variability, with the former still playing a more substantial role.

In the following sections, we investigate in more detail the origin of this feedback-driven variability.


\begin{figure}
        \includegraphics[width=\columnwidth]{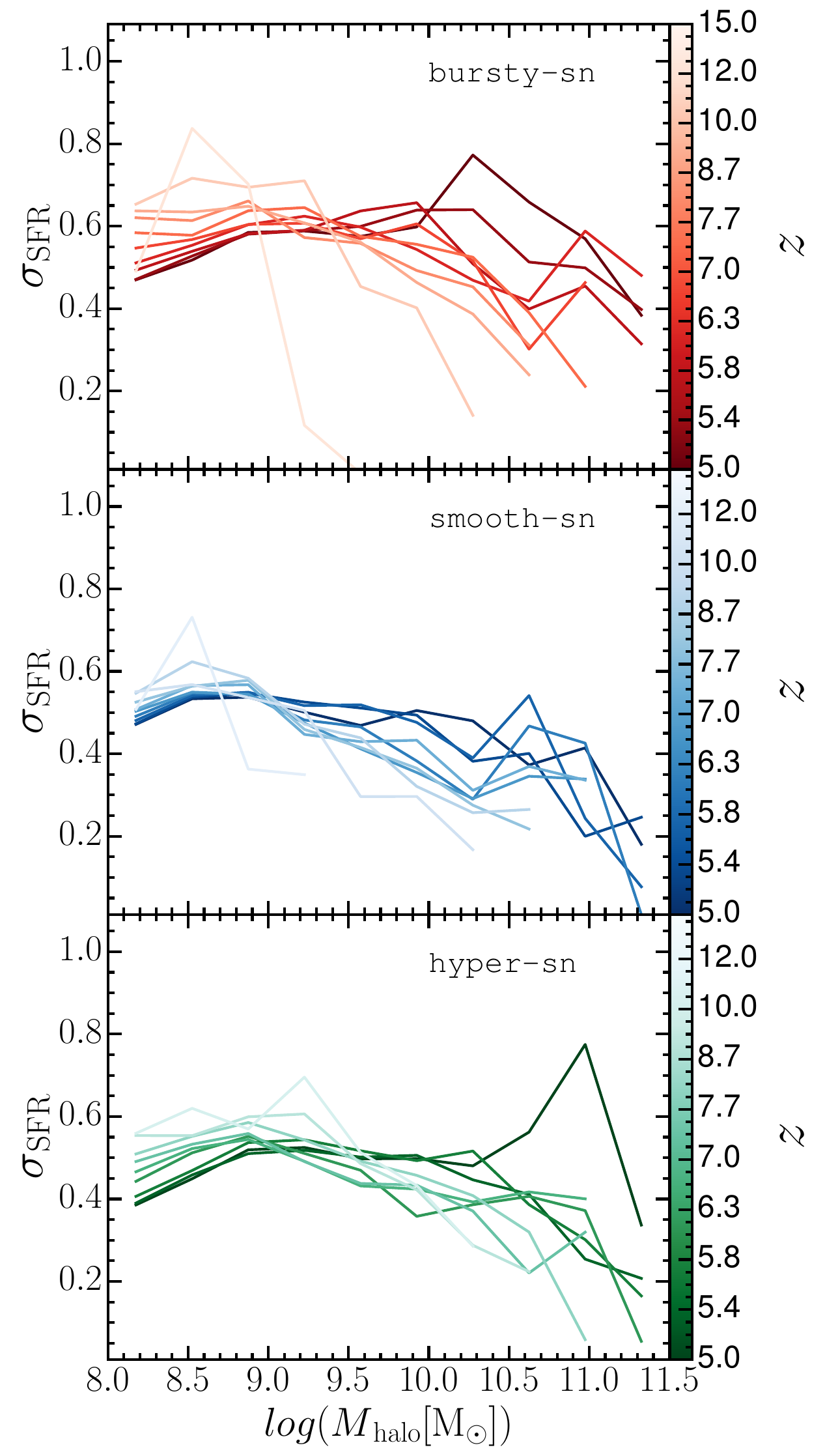}
  \caption{$\sigma\rm{_{SFR}}$ as a function of halo mass $M\rm{_{halo}}$ at different redshifts (indicated by the colorbar) for \texttt{bursty-sn} ({\it top panel}), \texttt{smooth-sn} ({\it middle}) and \texttt{hyper-sn} ({\it bottom}). }
  \label{fig:sigmasfr_mhalo_bins}
\end{figure}

\begin{figure}
        \includegraphics[width=\columnwidth]{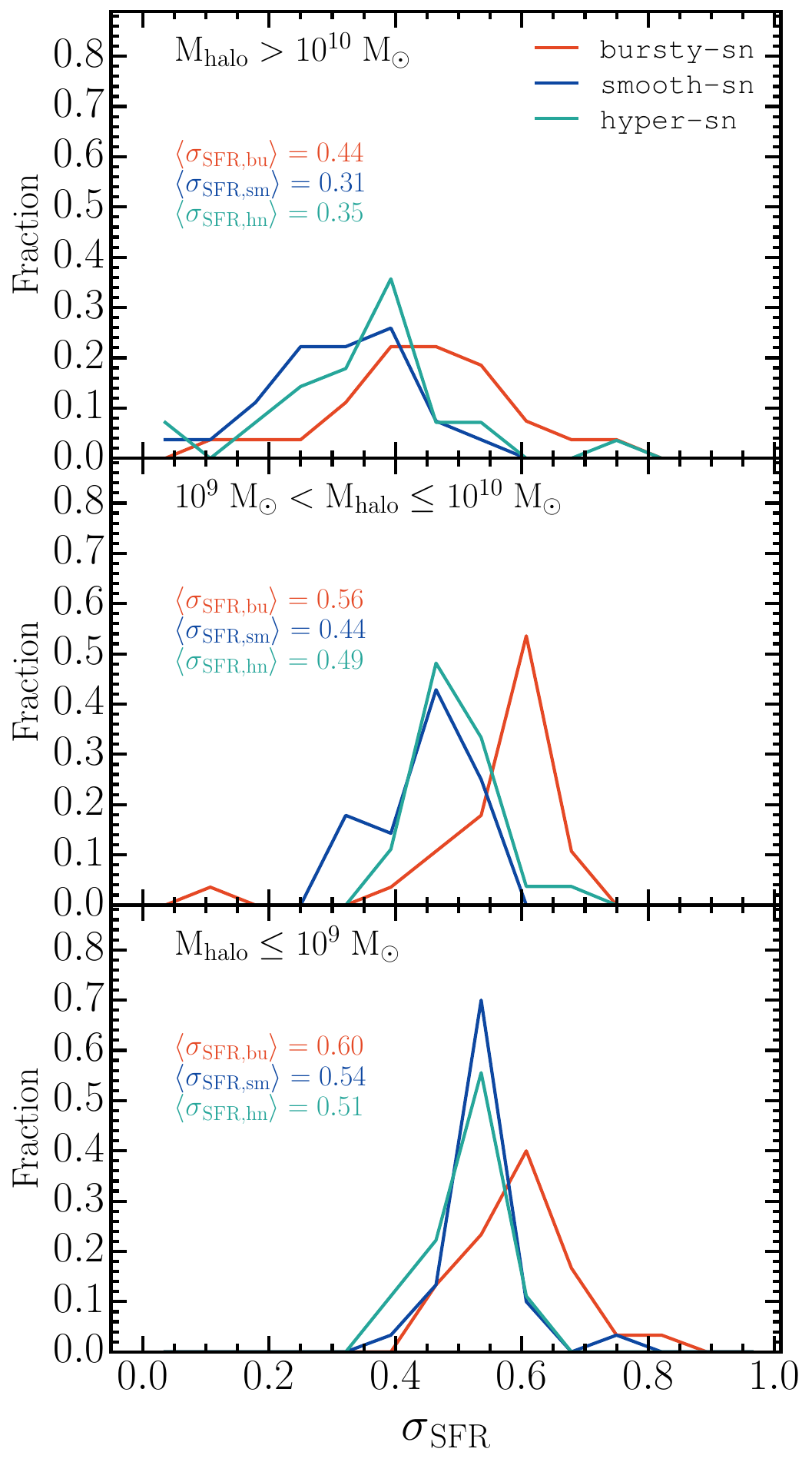}
  \caption{Distribution of $\sigma\rm{_{SFR}}$ in different $M\rm{_{halo}}$ bins for the three SN feedback models, as indicated by the colors. Numbers refer to the corresponding average standard deviation.}
  \label{fig:sfr_dist}
\end{figure}

\begin{figure}
        \includegraphics[width=\columnwidth]{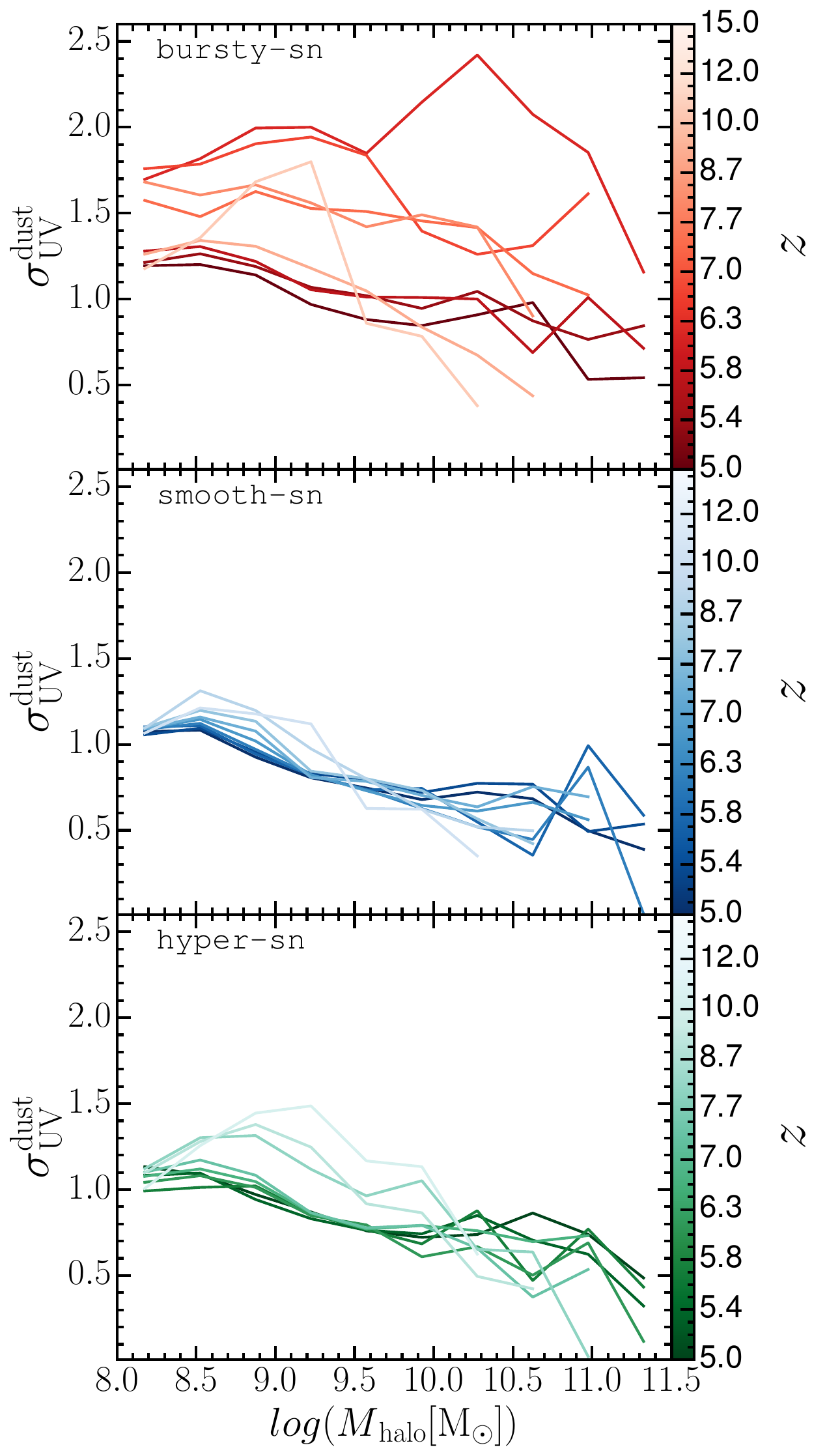}
  \caption{$\sigma\rm{_{UV}^{dust}}$ as a function of halo mass $M\rm{_{halo}}$ at different redshifts (indicated by the colorbar) for \texttt{bursty-sn} ({\it top panel}), \texttt{smooth-sn} ({\it middle}) and \texttt{hyper-sn} ({\it bottom}).}
  \label{fig:sigmauv_mhalo_bins}
\end{figure}

\begin{figure}
        \includegraphics[width=\columnwidth]{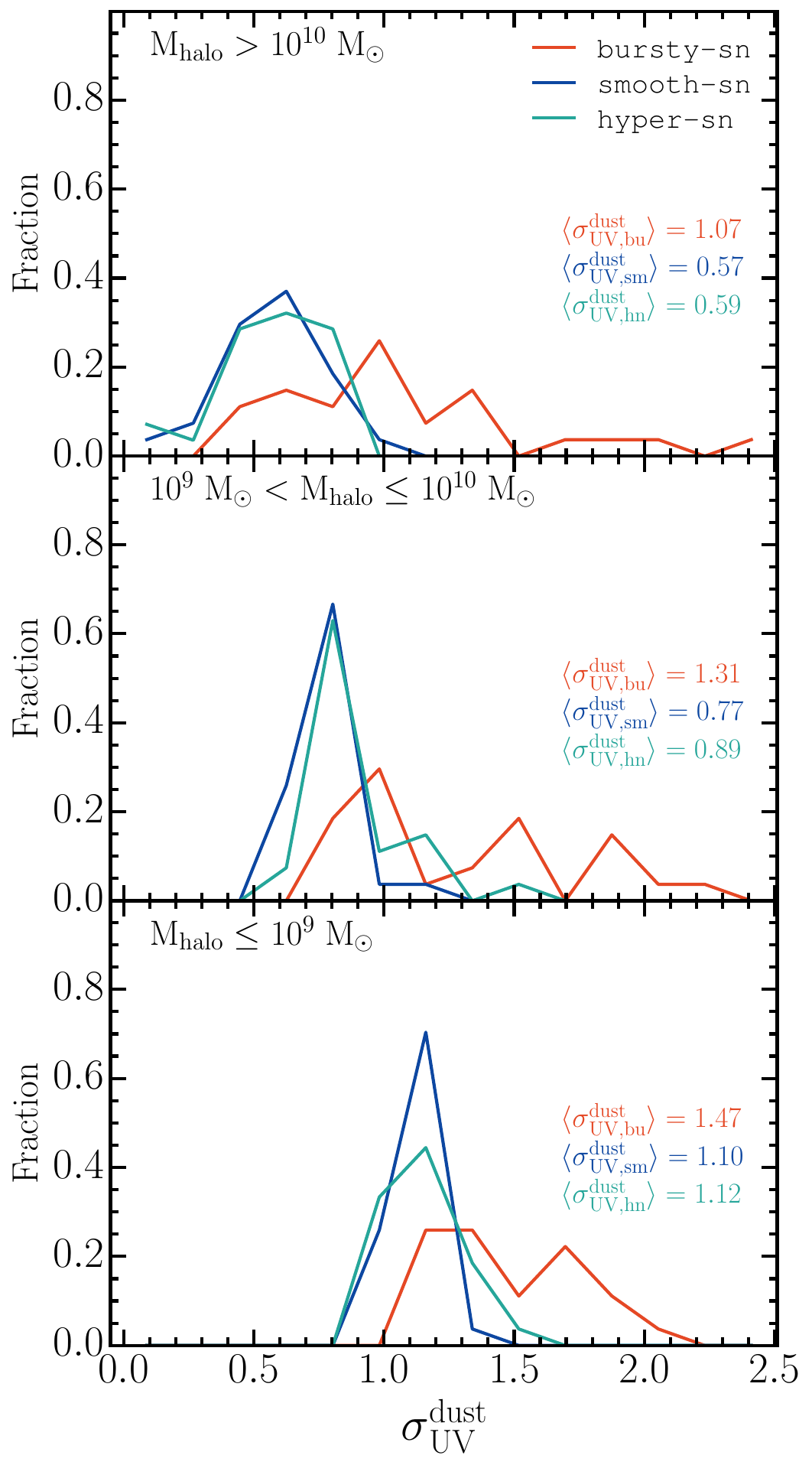}
  \caption{Distribution of $\sigma\rm{_{UV}^{dust}}$ in different $M\rm{_{halo}}$ bins for the three SN feedback models, as indicated by the colors. Numbers refer to the corresponding average standard deviation.}
  \label{fig:sigmauv_dist}
\end{figure}

\begin{figure}
        \includegraphics[width=\columnwidth]{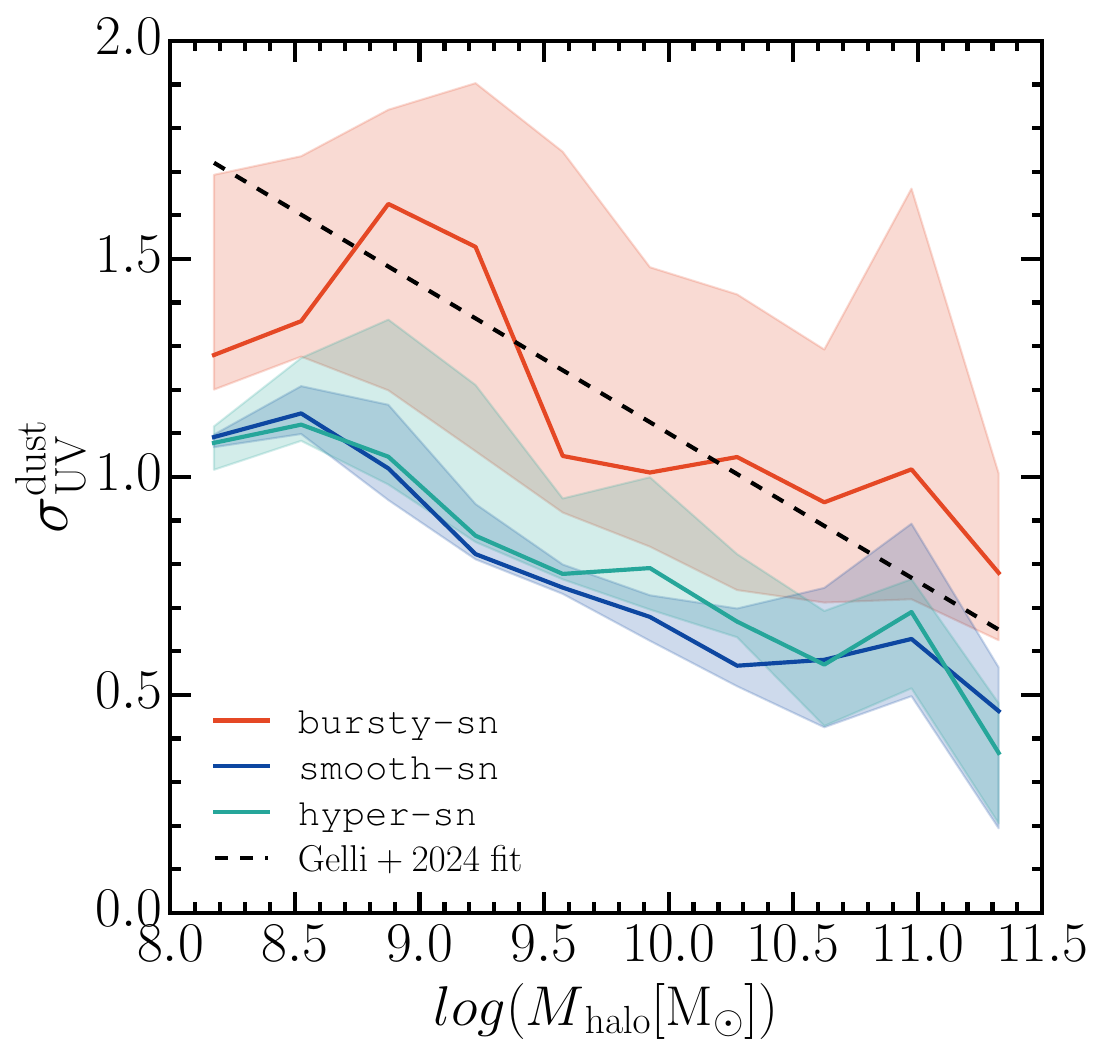}
  \caption{Dependence of $\sigma\rm{_{UV}^{dust}}$ on the DM halo mass $M\rm{_{halo}}$. The solid curves refer to the median value, while the shaded regions are the standard deviation. Colors refer to the different SN feedback models, while the black dashed curve represents the fit from \citet{gelli2024}.  }
  \label{fig:sigmauv_mhalo}
\end{figure}

\begin{figure*}
        \includegraphics[width=178mm]{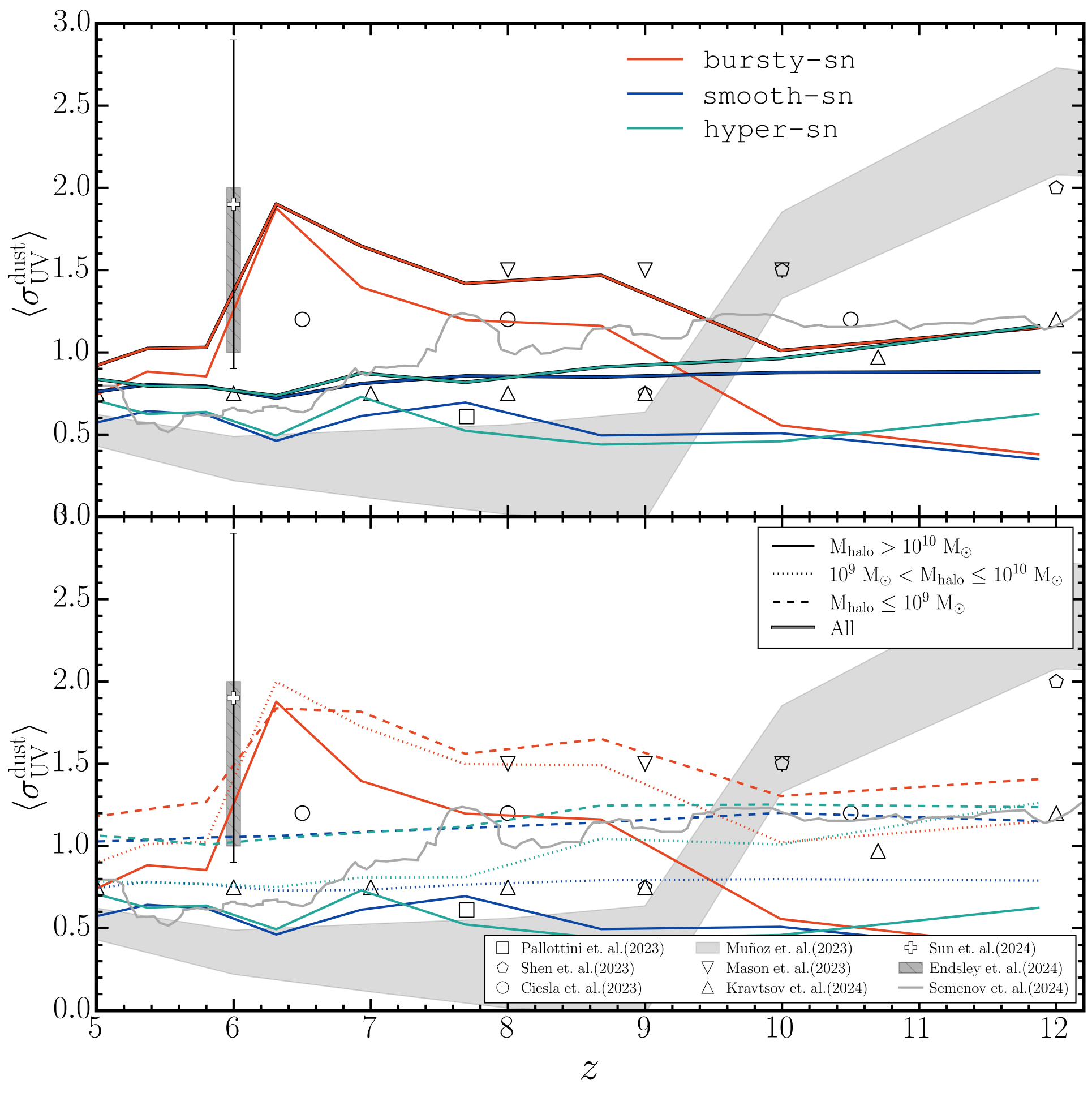}
  \caption{Redshift evolution of $\langle \sigma\rm{_{UV}^{dust}} \rangle$ for different halo mass bins i.e. $M\rm{_{halo}}>10^{10}$~M$_\odot$ (solid lines) (\textit{top panel}); $M\rm{_{halo}}<10^9$~M$_\odot$ (dashed) and $10^9$~M$_\odot< M\rm{_{halo}}<10^{10}$~M$_\odot$ (dotted) (\textit{bottom}). The curves for the entire sample are shown in the \textit{top panel} as solid lines with black edges. Colors refer to the three SN feedback models, while the values from observational and other theoretical studies \citep{pallottini_ferrera2023,Shen2023,ciesla2023,munoz2023,mason2023,kravtsov2024,sun2024,endsley2024,semenov2024} are shown in grey data points and shaded regions in both the panels.
  }
  \label{fig:sigmauv_evol}
\end{figure*}

\subsection{Star formation rate variability}
\label{sfr_ind}
In this section, we explore how different SN feedback models impact the variability of the SFR. Figure \ref{fig:sfr_individual} shows the star formation history of the most massive halo from \texttt{SPICE} at $z = 5$ for the three models. 
We note that at this redshift the virial mass of the halo is $2.7 \times 10^{11} \rm{\mathrm{M_{\odot}}}$ 
and the total stellar mass is
$2.1 \times 10^{9} \rm{\mathrm{M_{\odot}}}$, $1.9 \times 10^{10} \rm{\mathrm{M_{\odot}}}$ and $1.7 \times 10^{10} \rm{\mathrm{M_{\odot}}}$ in the \texttt{bursty-sn}, \texttt{smooth-sn} and \texttt{hyper-sn} model, respectively. 
It is evident that overall \texttt{bursty-sn} produces the largest fluctuations in the SFR, with deviations of up to two orders of magnitude. In contrast, the \texttt{smooth-sn} model shows smaller fluctuations.
In \texttt{hyper-sn}, the SFR is the lowest among the models
during the first 200 Myr (i.e. for lookback times larger than 700 Myr), when the impact of HN explosions is the strongest. However, at later times the SFR is similar to that of \texttt{smooth-sn}.
The fluctuations in \texttt{hyper-sn} are comparable to those in \texttt{smooth-sn}.

For a more quantitative investigation of the SFR variability, we first compute the median star-formation main sequence in different halo mass bins and redshift intervals (corresponding to bins of 100 Myr\footnote{This has been calculated from the SFR values of all the halos present in each 100 Myr time intervals and then computing the median SFR as a function of halo mass. We have also performed the same analysis using bin sizes of 30 and 50 Myr, and found no qualitative differences in our conclusions.}), SFR$_{\rm median}$.
In the inset of Figure \ref{fig:sfr_individual} we show the variability of the SFR for the same halo, defined as $\rm{\mathit{\delta}_{SFR}}={\rm log}_{10}\rm{(SFR/SFR_{median})}$, which quantifies how the SFR of the halo deviates from the median SFR computed over the entire sample of halos in the same mass bin.
In the inset of Figure~\ref{fig:sfr_individual} we show the distribution of $\delta_{\rm SFR}$ for this halo during its entire lifetime, together with the corresponding standard deviation, $\sigma_{\rm SFR}$. As expected, \texttt{bursty-sn} shows the widest distribution, with $\sigma_{\rm SFR,bu}=0.34$, which is more than double the one of \texttt{smooth-sn} with $\sigma_{\rm SFR,sm}=0.12$. \texttt{hyper-sn} produces a $\delta_{\rm SFR}$ distribution which is similar to the one from \texttt{smooth-sn}, with a slightly higher standard deviation of $\sigma_{\rm SFR,hn} = 0.14$, reflecting the stronger fluctuations present at $z>9$. \footnote{This behaviour persists when comparing objects of similar stellar and halo masses across the models with \texttt{bursty-sn} continuing to produce the largest scatter.} 
The level of variability could, in principle, depend on the size of the simulated volume, decreasing as fluctuations from individual objects partially cancel out in larger boxes. To test that, we repeated the analysis on smaller subvolumes (i.e. by selecting objects within cropped regions of the full box) and found out the variability remains essentially unchanged across different portions of the simulation, indicating that the box-size effect is minimal in this case. Larger boxes could still include higher overdensities and physical processes such as black-hole feedback, which might change the variability, but these are not included in the current \texttt{SPICE} simulations.

In Figure \ref{fig:sigmasfr_mhalo_bins} we examine the mass dependence of $\sigma_{\rm SFR}$ at different redshifts.
Typically, smaller halos (${M_{\rm halo}} < 10^{9}$~ M$_\odot$) exhibit a higher degree of variability as compared to more massive ones, suggesting a stronger impact of SN feedback at lower masses.
For halos with ${M_{\rm halo}} < 10^{9}$~ M$_\odot$, all three models have a similar redshift-dependent trend, with the variability decreasing at lower redshifts. 
However, the \texttt{bursty-sn} model consistently shows the highest variability in most redshift bins, whereas \texttt{smooth-sn} and \texttt{hyper-sn} have a similar behaviour, with the latter showing a slightly higher variability. At $z<6$, all the models become comparable, although \texttt{hyper-sn} shows the lowest $\sigma_{\rm SFR}$ values.
For ${M_{\rm halo}} > 10^{10}$~M$_\odot$, the trend is reversed, i.e. the SFR variability increases as redshift decreases. In \texttt{bursty-sn},  $\sigma_{\rm SFR}$ remains consistently high and shows little dependence on halo mass at lower redshifts. The behavior of \texttt{hyper-sn}  mostly mirrors the one of \texttt{smooth-sn} except the peak at $z\approx5$ for halos with masses around $10^{10.5} \rm{M_\odot}$, which reaches the highest SFR variability among all models.

To get an overview of the statistical behaviour of $\sigma_{\rm SFR}$, in Figure \ref{fig:sfr_dist} we present the distribution of $\sigma_{\rm SFR}$ in three halo mass bins: $M_{\rm halo} > 10^{10} {\rm M}_\odot$, $10^{9} {\rm M}_\odot < M_{\rm halo} \leq 10^{10} {\rm M}_\odot$, and $M_{\rm halo} \leq 10^{9} {\rm M}_\odot$ for the entire redshift range covered in Figure \ref{fig:sigmasfr_mhalo_bins}. In the \texttt{bursty-sn} model, strong and regular fluctuations  dominate the overall population, resulting in a $\sigma\rm{_{SFR}}$ distribution skewed towards higher values at all masses in comparison to \texttt{smooth-sn}.
The \texttt{hyper-sn} model produces distributions that typically fall between those of the other two models, although they are much more similar to those of \texttt{smooth-sn}. In the highest mass bin, though, \texttt{hyper-sn} exhibits a $\sigma\rm{_{SFR}}$ even higher than in \texttt{bursty-sn}, which aligns with the pronounced peak observed in Figure \ref{fig:sigmasfr_mhalo_bins}. 
For all models, the $\sigma_{\rm SFR}$ distributions are wider in the highest mass bin, indicating a broader range of star formation variability. As we move to lower mass halos, the distributions become progressively narrower, indicating a smaller range of variability but centered around higher $\sigma_{\rm SFR}$ values. Additionally, as noted earlier in Figure \ref{fig:sigmasfr_mhalo_bins}, the distributions for all models systematically shift towards higher $\sigma_{\rm SFR}$ values as halo mass decreases.

In the following section, we will explore how this mass and redshift-dependent SFR variability influences the variability of the UVLF.

\subsection{Variability of the  UV luminosity function}

The variability in the UVLF is shaped by three primary factors: the accretion history of galaxies, feedback-regulated star formation histories, and dust attenuation (see \citealt{Shen2023}). The combined effect of non-linear feedback and accretion history on the SFR is encapsulated in the term $\sigma_{\mathrm{SFR}}$, which has been introduced in the previous section. Interstellar medium (ISM) properties \citep{heckman2001,ciardi2002,alexandroff2015}, feedback-driven dust destruction and dust distribution within galaxies \citep{Aoyama2018,Ocvirk2024,Esmarian2024,Zhao2024} further influence the escape of UV photons, affecting the observed UVLF.

To examine the variability of the UVLF in \texttt{SPICE}, we compute the scatter in UV luminosity around the median UVLF (see Section~\ref{section:3.1}).
For a consistent comparison with previous studies, we discuss the variability of the dust-attenuated UVLF ($\rm{\mathit{\sigma}_{UV}^{dust}}$), which generally shows a variability higher than the one of the intrinsic one \citep{Shen2023,pallottini_ferrera2023}.  In \texttt{SPICE}, the UVLF variability is influenced by both intrinsic flux fluctuations and dust attenuation, with the former playing a more dominant role. However, in approximately $25-35$$\%$ of cases (with the \texttt{smooth-sn} model showing the highest fraction), dust attenuation slightly reduces the overall variability of $\lesssim 4-5\%$ with respect to the intrinsic scatter.

Figure \ref{fig:sigmauv_mhalo_bins} shows the scatter in UV luminosity, $\sigma_{\rm UV}^{\rm dust}$, as a function of $M_{\rm halo}$ in the range $z=5-15$. At the highest redshifts, all models exhibit a similar dependence on halo mass, with smaller halos showing a larger UVLF variability, likely due to efficient SN feedback in this mass regime \citep{gelli2024}. As expected, \texttt{bursty-sn} has the highest variability among all models, with a UV luminosity scatter which starts to rise at $z\lesssim 10$ and reaches $\sigma_{\rm UV}^{\rm dust} \approx 2.5$ for $M_{\rm halo} \approx 10^{10.3}$~M$_{\odot}$ at $z \approx 6.5$. At lower redshift, $\sigma_{\rm UV}^{\rm dust}$ decreases again.
In comparison, for \texttt{smooth-sn} the UV luminosity scatter is always below $\sigma_{\rm UV}^{\rm dust}\approx 1.3$, whereas \texttt{hyper-sn} shows slightly higher variability, with a maximum value of $\sigma_{\rm UV}^{\rm dust} \approx 1.5$.
Indeed, \texttt{hyper-sn} exhibits a variability comparable to the one in \texttt{bursty-sn} until $z \approx 9$, while at lower redshift the variability decreases as a result of the reduced fraction of HN,
and becomes more similar to the \texttt{smooth-sn} one. The same behaviour had already been observed in Figure \ref{fig:sfr_individual} with respect to the SFR. At $z < 6$ all models have nearly identical variability levels, except for \texttt{bursty-sn} which shows larger fluctuations, with a maximum deviation up to 0.3-0.4 mag.

Similarly to Figure~\ref{fig:sfr_dist}, in Figure \ref{fig:sigmauv_dist} we show the distribution of $\rm{\mathit{\sigma}_{UV}^{dust}}$ in different halo mass bins. As found for the SFR, the curves are shifted towards higher $\rm{\mathit{\sigma}_{UV}^{dust}}$ for smaller halo masses, with \texttt{bursty-sn} having a broader distribution and the highest peak value overall. \texttt{smooth-sn} and \texttt{hyper-sn} show  distributions similar to each other, with the exception of the intermediate mass range (i.e. $10^{9} {\rm M}_\odot < M_{\rm halo} \leq 10^{10} {\rm M}_\odot$), where the latter is somewhat wider. In this mass regime, the average $\rm{\mathit{\sigma}_{UV}^{dust}}$ is about 0.89 for \texttt{hyper-sn} and 0.77 for \texttt{smooth-sn}, whereas \texttt{bursty-sn} has a higher average of 1.31.
For \texttt{bursty-sn}, the average $\rm{\mathit{\sigma}_{UV}^{dust}}$ increases with decreasing mass bin, ranging from 1.07 to 1.47, with a significant fraction of objects having $\rm{\mathit{\sigma}_{UV}^{dust}} > 1.5$. The maximum value reached in \texttt{smooth-sn} is $\sigma_{\rm UV}^{\rm dust} = 1.5$, with the average $\rm{\mathit{\sigma}_{UV}^{dust}}$ increasing from $0.57$ to $1.1$ from higher to lower halo mass bins.
The \texttt{hyper-sn} model falls in-between the other two with average $\rm{\mathit{\sigma}_{UV}^{dust}}$ values ranging from $0.59$ to $1.12$.

In Figure \ref{fig:sigmauv_mhalo} we show the median $\sigma\rm{_{UV}^{dust}}$ as a function of $M\rm{_{halo}}$, where the former is calculated as the median value in each individual mass bin from all the curves corresponding to different redshift bins in Figure~\ref{fig:sigmauv_mhalo_bins}. 
We find that the median curves in all models exhibit a similar slope, confirming that lower mass halos are more sensitive to feedback effects, producing more fluctuations compared to massive halos. This is primarily due to the shallower potential wells of lower mass halos, which facilitate repeated cycles of inflow, star formation, and outflow \citep{gelli2020,stern2021,furlanetto2022,legrand2022,gurvich2023,byrne2023,hopkins2023}. Among the models,  \texttt{bursty-sn}  consistently produces the highest values. Additionally, the scatter around the median differs among the models, and, as expected, the \texttt{bursty-sn} model shows the largest scatter because of the wider range of variability in UVLF, while \texttt{smooth-sn} has the smallest. 
We also note that the slope of our curves is consistent with the one from the analytical fit by \citet{gelli2024} based on the results of the \texttt{FIRE-2} simulation at $z \approx 8$ \citep{sun2023}, although the amplitude is matched only by the \texttt{bursty-sn} model.

Finally, in Figure \ref{fig:sigmauv_evol} we show the redshift evolution of the average UV luminosity scatter $\langle \sigma_{\rm UV}^{\rm dust} \rangle$ in different halo mass bins (at all redshifts), as well as computed over the full sample.
We observe that in all models  $\langle \sigma_{\rm UV}^{\rm dust} \rangle$ is largest for smaller halos, consistently with Figure \ref{fig:sigmauv_mhalo}.
The curves for both the \texttt{smooth-sn} and \texttt{hyper-sn} models exhibit a clear and consistent trend across redshifts, with $\langle \sigma_{\rm UV}^{\rm dust} \rangle$ generally decreasing with decreasing redshift with an exception at the highest halo mass bin where the scatter in UV luminosity remains roughly constant. 
In contrast, across all masses and redshifts, the \texttt{bursty-sn} model consistently exhibits the highest $\langle \sigma_{\rm UV}^{\rm dust} \rangle$, indicating more pronounced fluctuations in UVLF compared to the other models.
Similarly to the trend observed in Figure \ref{fig:sigmauv_mhalo_bins}, also here we see that in the \texttt{bursty-sn} model, as the UVLF variability increases rapidly below $z\approx10$, has a peak of $\langle \sigma_{\rm UV}^{\rm dust} \rangle \approx 2$ at $z \approx 6.3$, and then it drops sharply. 
Since the halo assembly history is similar in all models, this behaviour is unlikely to be driven solely by halo mass evolution. For example, its origin may also be connected to galaxy morphology, gas accretion history, level of turbulence in the ISM, and the rise of a UVB, as around this time the reionization process begins to accelerate for the \texttt{bursty-sn} model. This will be addressed in more detail in a future study.
Notably, at $z \gtrsim 10$ for halos with $M_{\rm halo}>\rm{10^{9} M_\odot}$, \texttt{hyper-sn} produces the highest variability among all three models because of the impact of the stronger HN explosions dominating at that time.

We compare the results from the \texttt{SPICE} simulations to values extracted from observational \citep{ciesla2023,endsley2024}, as well as theoretical \citep{munoz2023,mason2023,Shen2023,pallottini_ferrera2023,sun2023,sun2023a,sun2024,kravtsov2024,semenov2024} studies.
Employing a semi-empirical approach, \citet{munoz2023} found a value of $\langle \sigma\rm{_{UV}^{\rm dust}} \rangle$ as high as 2.5 at $z\approx 12$, which drops to 0.8 at $z \leq 10$. While none of our models reproduces such large high-$z$ values, at lower redshift there is a better agreement, although in this case our $\langle \sigma\rm{_{UV}^{\rm dust}} \rangle$ are always slightly larger.
\citet{mason2023} utilized a fully empirical framework to predict that a $\langle \sigma\rm{_{UV}^{\rm dust}} \rangle$ up to 1.5 in the range $8<z<10$ is necessary for models to align with existing observations. These numbers are similar to those predicted in our {\tt bursty-sn} model.
Following a similar methodology, \citet{Shen2023} found a lower value of 0.75 at $z \approx 9$, which is instead consistent with those predicted by the {\tt smooth-sn} and {\tt hyper-sn} models. At $z \approx 10$ and 12, though, they predict values of 1.5 and 2.0, respectively, which are higher than those of our models.
Using a semi-analytical approach, \citet{sun2024} (but see also \citealt{sun2023a,sun2023}, where bursty star formation in \texttt{FIRE-2} simulation has been used to explain high redshift \texttt{JWST} observations) found the most probable UVLF variability value to be $\approx1.9\pm 1.0$ at $z\approx6$. At the same redshift, \citet{endsley2024} found a variability in the range 1-2, obtained by analyzing Lyman-break galaxies  assembled from ACS+NIRCam imaging in the GOODS and Abell 2744 fields. These values are generally reproduced in \texttt{bursty-sn}, as well as by halos with $M_{\rm halo}\leq\rm{10^{9} M_\odot}$ in the other models.
\citet{ciesla2023}, using spectral energy distribution modeling with the \texttt{CIGALE} code, and assuming a non-parametric star formation history on the \texttt{JADES} public catalog, found $\langle \sigma\rm{_{UV}^{\rm dust}}\rangle \sim 1.2$ in the range $6.5 < z < 10.5$, which is fairly consistent with our {\tt bursty-sn} model. 
\citet{pallottini_ferrera2023} analyzed 245 galaxies from the \texttt{SERRA} simulation suite at $z\approx 7.7$, finding a $\langle \sigma\rm{_{UV}^{\rm dust}} \rangle$ of 0.61. We note, though, that the method employed for this evaluation differs from that used in our study and other works (see appendix \ref{Appendix1} for more details). 
Finally, \citet{semenov2024} investigated the importance of turbulent star formation by utilizing detailed modelling of cold turbulent ISM, star formation and feedback through zoom-in high resolution simulation of an early forming Milky Way analog, and they found that the variability decreases with time, similarly to the predictions of our models, with the exception of the $6<z<9$ range in \texttt{bursty-sn}. It is also interesting to note that the curves for the \texttt{smooth-sn} and \texttt{hyper-sn} models in the lowest mass bin aligns very well with \citet{semenov2024} till $z\approx7.5$, whereas after that the latter shows much lower values. 

In general, we can conclude that the constraints from observational and other theoretical models can not specifically distinguish between the SN feedback models in \texttt{SPICE} due to their large scatter in $\sigma\rm{_{UV}^{\rm dust}}$ across different redshifts.

\section{Conclusions}
\label{section:4}

In this study we investigate the role played by SN feedback in driving the variability of the UVLF using the suite of radiation-hydrodynamic simulations \texttt{SPICE}.
This includes three models for SN feedback, which differ in terms of explosion energy and timing. 
The \texttt{bursty-sn} model, characterized by intense and episodic supernova explosions, shows the highest SFR variability, leading to significant fluctuations in UV luminosity. 
The \texttt{smooth-sn} model, in which energy injection happens more continuously and thus the effect of feedback is less disruptive, produces a higher and less variable SFR, resulting also in a lower UVLF variability. 
The \texttt{hyper-sn} model shows an intermediate trend, with higher variability at early times due to HN effects, and a behaviour similar to the one of \texttt{smooth-sn} at lower redshifts, when the fraction of HN events decreases. Our main findings can be summarized as follows:

\begin{itemize}

    \item The good agreement between all models and the observed UV luminosity functions discussed in \citet{Bhagwat2024} is obtained even at the highest redshifts analyzed here, i.e. $z\approx 11$ and 12. 
    
    \item We find that not only the most massive objects contribute to the bright end of the LF, i.e. with dust corrected magnitude $\rm{\mathit{M}_{UV}^{dust}} \leq -17$, but also those with $\rm{\mathit{M}_{*} \sim 10^{(6.5-7)} M_{\odot}}$ can be a very bright object detected by \texttt{JWST}. The median stellar mass of these bright objects in our models is consistent with those derived by \citet{mason2015,mason2023} at all redshifts.

    \item The nature of SN feedback has a significant impact on the temporal evolution of the UVLF, as we show for time intervals of 10, 30, and 50 Myr at $z\approx10$ and 8. The \texttt{bursty-sn} model produces the highest UVLF fluctuations, with a deviation of more than 1 dex. In comparison, the evolution of the UVLF in \texttt{smooth-sn} and \texttt{hyper-sn} is more gradual, with the latter having much higher fluctuations than the former at $z\approx10$. 

    \item The burstiness of star formation and UVLF is strongly influenced by the SN feedback, with the \texttt{bursty-sn} and \texttt{smooth-sn} models consistently exhibiting the highest and lowest variability, respectively. The disruptive nature of the \texttt{bursty-sn} model leads to both higher median values and larger scatter in $\sigma_{\rm UV}^{\rm dust}$, indicating stronger fluctuations in dust-obscured UV emission. In contrast, the \texttt{hyper-sn} model displays an intermediate behavior, balancing between the extremes of the other two models.

    \item All models show a similar dependence of the SFR and UVLF variability on the halo mass, with the extent of the fluctuations being higher for the lowest mass haloes, which are more susceptible to feedback effects. We find that the median $\sigma\rm{_{UV}^{dust}}$ as a function of $M\rm{_{halo}}$ in all models exhibit a similar slope, consistent with the one from the analytical fit from the \texttt{FIRE-2} simulation at $z \approx 8$ by \citet{sun2023}, although the amplitude is matched only by the \texttt{bursty-sn} model.

    \item The redshift dependence of the average standard deviation $\langle \sigma_{\rm UV}^{\rm dust} \rangle$ is similar in  \texttt{smooth-sn} and \texttt{hyper-sn}, with a variability which remains almost constant, although in the latter model it is slightly higher. In contrast, the \texttt{bursty-sn} model shows an increase in $\langle \sigma_{\rm UV}^{\rm dust} \rangle$ from $z \approx 10$ to $\approx 6$, before declining. Additionally, the \texttt{bursty-sn} model consistently exhibits the highest values of $\langle \sigma_{\rm UV}^{\rm dust} \rangle$ across all masses and redshifts.

     \item We achieve the maximum UVLF variability as $\sigma_{\rm UV}^{\rm dust} \sim 2.5$ in the {\tt bursty-sn} model. Conversely, the smoother star formation history typical of the {\tt smooth-sn} model fails to induce a large variability, with values of $\sigma_{\rm UV}^{\rm dust}$ below 1.3. The {\tt hyper-sn} model lies is in between the other two, although its characteristics are more similar to those of the {\tt smooth-sn} one, with $\sigma_{\rm UV}^{\rm dust}$ extending up to 1.5.
\end{itemize}

This work emphasizes the critical role of SN feedback in shaping the variability of the UVLF. 
While our study suggests that SN feedback is the primary mechanism influencing the UVLF variability, the observed features arise from a complex interplay of multiple factors (i.e. UVB, gas accretion history, ISM turbulence) that requires further detailed investigation.
Here, though, we provide new insights into how variations in feedback strength and timing impact the burstiness of star formation and UV emission, and we emphasize the importance of explaining how the variability depends on both mass and redshift, an aspect often overlooked in prior studies. Our investigation also suggests the UVLF variability may alleviate the bright galaxy tension observed by JWST at high redshifts.

\section*{Acknowledgements}
All the analysis were carried out on the machines of Max Planck Institute for Astrophysics (MPA) and Max Planck Computing and Data Facility (MPCDF). We thank the anonymous reviewer and the editor for the useful comments which helped to improve the manuscript. AB thanks the entire EoR research group of MPA for all the encouraging comments for this project. This work made extensive use of publicly available software packages : \texttt{numpy} \citep{vander2011}, \texttt{matplotlib} \citep{Hunter2007}, \texttt{scipy} \citep{Jones2001}. Authors thank the developers of these packages.

\section*{Data Availability}
The final data products from this study will be shared on reasonable request to the authors.



\bibliographystyle{mnras}
\bibliography{mnras} 




\appendix
\section{Comparison between different methods for computing $\sigma\rm{_{SFR}}$}
\label{Appendix1}

\begin{figure}
        \includegraphics[width=\columnwidth]{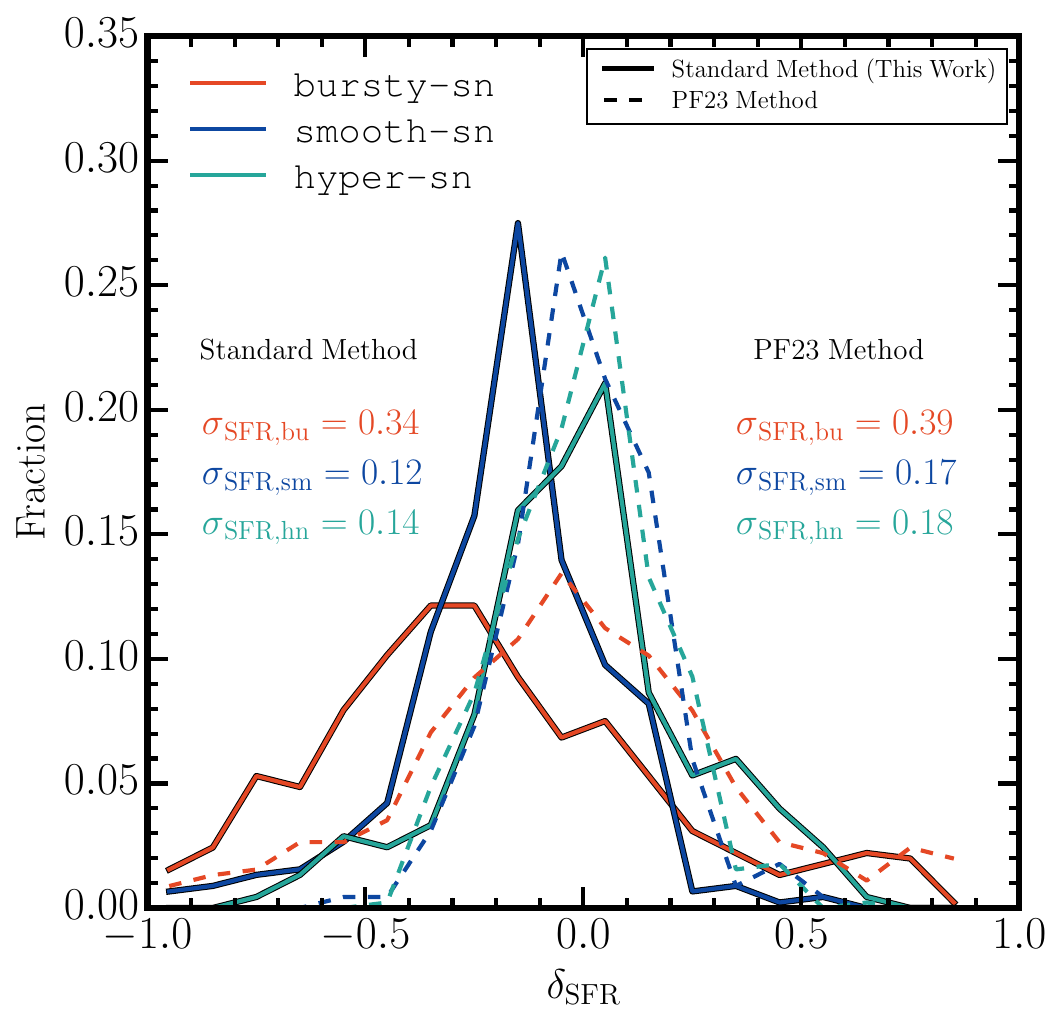}
  \caption{Distribution of $\delta_{\rm{SFR}}$ computed from the temporal evolution of the most massive halo at $z=5$ as shown in Figure \ref{fig:sfr_individual}. Colors refer to different SN feedback models. The solid and dashed curves represent the results obtained using the standard method and those from the method proposed by \citet{pallottini_ferrera2023}. Numbers refer to the corresponding standard deviation.}
  \label{fig:deltasfr_compare}
\end{figure}

\begin{figure}
        \includegraphics[width=\columnwidth]{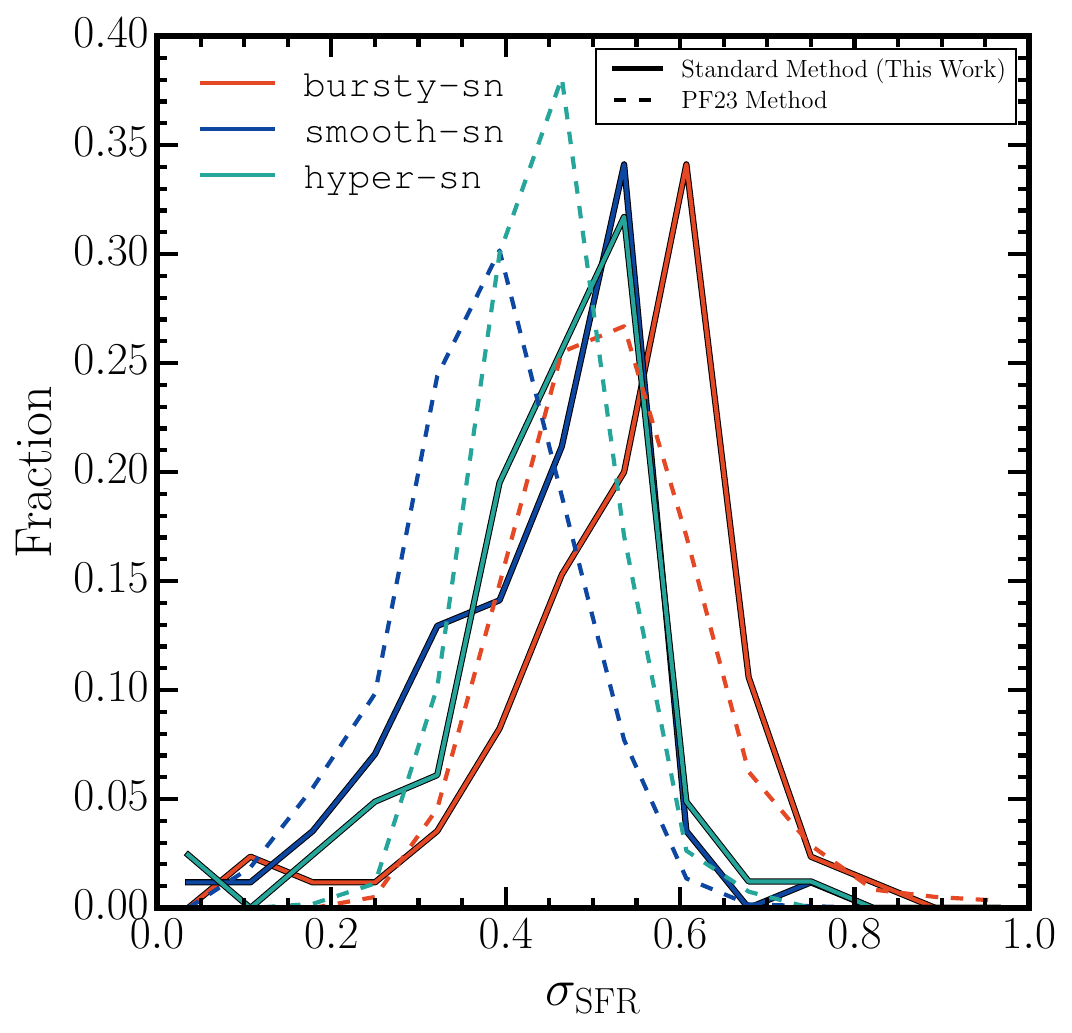}
  \caption{Distribution of $\sigma_{\rm{SFR}}$ computed over the entire sample of halos present at $z=5$. Colors refer to different SN feedback models. Curves represent the two different methods as in Figure \ref{fig:deltasfr_compare}.}
  \label{fig:sigmasfr_compare}
\end{figure}
In this section, we discuss how our major findings are affected by adopting the method proposed by \citet{pallottini_ferrera2023} (hereafter referred to as \texttt{PF23}) in the calculation of the variability. We begin by analyzing the temporal evolution of the SFR for the most massive halo at $z=5$  illustrated in Figure \ref{fig:sfr_individual}. Each curve is fitted with a third-order polynomial on a log-linear scale (i.e. $\text{SFR}_{\text{fit}}$), following the technique used by \texttt{PF23}. The stochastic variability of the SFR is then quantified by computing the residuals, defined as $\delta_{\rm{SFR}} = \log_{10}(\text{SFR}/\text{SFR}_{\text{fit}})$, from which we derive the standard deviation, $\rm{\mathit{\sigma}_{SFR}}$. The results of this approach are compared with the standard method presented in Section \ref{sfr_ind} and summarized in Figure \ref{fig:deltasfr_compare}. We observe that the $\delta_{\rm{SFR}}$ distribution using the \texttt{PF23} method is slightly shifted toward higher values across all models compared to the standard method. This shift is reflected in the slightly higher values of $\rm{\mathit{\sigma}_{SFR}}$, with 0.39 for \texttt{bursty-sn}, 0.17 for \texttt{smooth-sn}, and 0.19 for \texttt{hyper-sn}, each exceeding the standard method's results by approximately $\Delta(\mathit{\sigma}_{\rm{SFR}}) \sim 0.04-0.05$. To examine the behavior of the entire halo population at $z=5$, Figure \ref{fig:sigmasfr_compare} presents the distribution of $\sigma_{\mathrm{SFR}}$ calculated using both methods across all SN feedback models. Interestingly, we observe that the \texttt{PF23} method yields a distribution of $\sigma_{\mathrm{SFR}}$ values that are slightly lower across all models when compared to the standard method. This finding contrasts with the trend seen in Figure \ref{fig:deltasfr_compare}. Specifically, we note that the peak of the $\sigma_{\mathrm{SFR}}$ distribution shifts by approximately $0.15 - 0.2$.

The key difference lies in the \texttt{PF23} method’s focus on the temporal evolution of individual halos up to the redshift of interest, which emphasizes fluctuations around the median evaluated for the same halo, differently from other methods which calculate the median from the full sample of halos.  This distinction leads to a fundamentally different interpretation of the underlying variability being measured, with the \texttt{PF23} method offering a more detailed view of individual halo evolution over time.

\bsp	
\label{lastpage}
\end{document}